\documentclass[iop]{emulateapj}

\newcommand{\jcap}{JCAP}
\newcommand{\APh}{APh}
\renewcommand{\prd}{PhRvD}
\renewcommand{\apjl}{ApJL}

\renewcommand{\theta}{\vartheta}
\newcommand{\Mpc}{\,Mpc$^{-3}$}

\newcommand{\yr}{\,yr$^{-1}$}
\newcommand{\s}{\,s$^{-1}$}
\newcommand{\ergs}{\mbox{\,erg\,\s}}

\usepackage{graphicx}% Include figure files
\usepackage[colorlinks,urlcolor=blue,citecolor=blue,linkcolor=blue]{hyperref}
\usepackage{amsmath}
\usepackage{bm}

\shorttitle{``Espresso" Acceleration of UHECRs}
\shortauthors{Caprioli}

\begin{document}

%%\preprint{}

\title{``Espresso'' Acceleration of Ultra-High-Energy Cosmic Rays}
\author{Damiano Caprioli}
\affil{Department of Astrophysical Sciences, Princeton University, 
    4 Ivy Ln., Princeton NJ 08544, USA}
\email{caprioli@astro.princeton.edu}

\begin{abstract}
We propose that ultra-high-energy (UHE) cosmic rays (CRs) above $10^{18}$\,eV  are produced in relativistic jets of powerful active galactic nuclei via an original mechanism, which we dub ``espresso" acceleration:
``seed" galactic CRs with energies $\lesssim 10^{17}$\,eV that penetrate the jet sideways receive a ``one-shot'' boost of a factor of $\sim\Gamma^2$ in energy, where $\Gamma$ is the Lorentz factor of the relativistic flow.
For typical jet parameters, a few percent of the CRs in the host galaxy can undergo this process, and powerful blazars with $\Gamma\gtrsim 30$ may accelerate UHECRs up to more than $10^{20}$\,eV.
The chemical composition of espresso-accelerated UHECRs is determined by that at the Galactic CR knee and is expected to be proton-dominated at $10^{18}$\,eV and increasingly heavy at higher energies, in agreement with recent observations made at the Pierre Auger Observatory.

\end{abstract}
                            % Classification Scheme.
%Use showkeys class option if keyword
%display desired

\section{THE ORIGIN OF ULTRA-HIGH--ENERGY (UHE)\\ COSMIC RAYS (CRS)} 
The mechanism responsible for the acceleration of highest-energy CRs detected at Earth is one of the most prominent open questions in astrophysics. 
Possible sources of particles as energetic as 10$^{20}$\,eV are limited to  gamma-ray bursts (GRBs) \cite[e.g.,][]{vietri95,waxman95}, newly born millisecond pulsars \cite[e.g.,][]{Blasi+00,Fang+12}, and active galactic nuclei (AGNs) \cite[e.g.,][]{ostrowsky00,Murase+12}, which can ---in principle--- generate the electric potential necessary to produce UHECRs \citep{hillas84,Dermer07}.
Recent improved measurements of  extensive atmospheric air showers induced by UHECRs are providing  unprecedented information on their chemical composition.
The Pierre Auger Observatory measured a composition compatible with pure protons around $10^{18}$\,eV, and increasingly heavy at higher energies \citep{Auger14a}. 
Such results are not inconsistent with Telescope Array's observations \citep{PAO-TA15}, which favor a lighter chemical composition, especially if interpreted in light of the LHC data \citep{Pierog13}.
The evidence for nuclei with charge $Z\gtrsim 1$ in the most energetic UHECRs may be of paramount importance for unraveling their origin.

In this Letter, we exploit some basic properties of particle acceleration in relativistic flows to outline a simple but general mechanism for the production of UHECRs in powerful AGNs.
We call such a process the \emph{``espresso'' mechanism}, since a very fast flow (the AGN jet) provides a \emph{one-shot} acceleration of energetic \emph{seeds} (galactic CRs). 
The implications of such a scenario for CR spectrum and chemical composition at energies  $E\gtrsim 10^{15}$\,eV are discussed and compared with observations. 

\section{AGNs as Potential Sources}
We briefly review the requirements for candidate UHECR sources (energetics, luminosity, and confinement) and check that AGNs satisfy all of them, if the highest-energy CRs are heavy nuclei.

The energy generation rate necessary to sustain the UHECR flux above $1{\rm EeV}$ is $\approx$$5.4\times 10^{45}$\,erg\Mpc\yr \citep[e.g.,][]{Katz+13}.
Typical bolometric AGN luminosities range between $L_{\rm bol}\approx$$10^{43}\ergs$ for Seyfert galaxies and radio-quiet quasars, to $L_{\rm bol}\gtrsim 10^{47}\ergs$ for radio-loud quasars \citep[e.g.,][]{woo-urry02};
since the typical density of active galaxies is $\approx10^{-4} $\Mpc, 10--20\% of which is radio-loud \citep{Jiang+07}, the total AGN energy generation rate (in photons) is $\approx$$5\times 10^{48}$\,erg\Mpc\yr. 
AGNs would meet the energy requirement if they emitted a fraction $\gtrsim 10^{-3}$ of their bolometric luminosity in UHECRs.

If the flow is expanding, the rapid acceleration up to an energy $E_{\rm max}$ may also require a minimum source luminosity \citep{Waxman04}:
\begin{equation}\label{eq:Lcr}
L_{*}\approx 5\times 10^{42}~\frac{\Gamma^2}{\beta}\left(\frac{E_{\rm max}}{Z_{26}\, 10^{20}{\rm \, eV}}\right)^2 \mbox{erg\s},
\end{equation}
where $\Gamma$ and $\beta c$ are the flow Lorentz factor and speed, and $Z_{26}$ is the particle charge in units of the iron nucleus charge.
For powerful AGNs, $L_{\rm bol}\gtrsim L_{*}$, provided that UHECRs with $E_{\rm max}\sim 10^{20}$\,eV are heavy nuclei and that $\Gamma\lesssim 50$.
However, $L_{\rm bol}$ represents a lower limit on the actual source power, and the luminosity in CRs can be much larger than that in photons: for instance, $L_{\rm CR}\approx 10^{41}\mbox{erg\s}\gg L_{\rm bol}\lesssim 10^{35}\mbox{erg\s}$ in Tycho's supernova remnant (SNR) \citep{tycho}. 
In blazars the jet power is inferred to be a factor of 10--100 larger than $L_{\rm bol}$ \cite[e.g.,][]{GTG09}.

The constraint in Equation \ref{eq:Lcr} is similar (though not equivalent) to the so-called Hillas criterion \citep{hillas84},
\begin{equation}\label{eq:hillas}
B_{\mu{\rm G}}R_{\rm kpc}\gtrsim \frac{4\beta}{Z_{26}}\frac{E_{\rm max}}{10^{20}{\rm eV}},
\end{equation}
which expresses the maximum energy achievable in a system  with  magnetic field $B$ (in $\mu$G) and size $R$ (in kpc). 
For typical AGN jets, $B_{\mu{\rm G}}R_{\rm kpc}\gtrsim 1$, allowing confinement and  acceleration of nuclei up to $E\gtrsim 10^{20}$\,eV.

These considerations support the hypothesis that AGNs can accelerate EeV protons and iron nuclei up to the highest observed energies even for rather small acceleration efficiencies ($\gtrsim 10^{-4}$ of the jet power).

\section{Acceleration in relativistic flows}
Non-relativistic supersonic flows that dissipate their energy into shocks can efficiently energize CRs via diffusive shock acceleration \citep[e.g.,][]{DSA}; 
in relativistic flows, instead, diffusive acceleration may be quite suppressed \citep[see, e.g.,][]{ss11}, and different mechanisms are needed to produce energetic particles.

Let us consider a relativistic flow with Lorentz factor $\Gamma$ and velocity $\beta c\,\bm{\hat{x}}$ in the laboratory frame, and a particle with initial energy $E_{\rm i}$ and momentum
\begin{equation}
 \bm{p}_{\rm i}\simeq E_{\rm i}(\mu_{\rm i},-\sqrt{1-\mu_{\rm i}^2},0),
\end{equation}
where $\mu\equiv p_x/|\bm{ p}|$; once in the flow, its energy in the flow frame is
\begin{equation}
\begin{split}
 E'_{\rm i} & = \Gamma (E_{\rm i}-\beta p_{{\rm i},x}) = \Gamma E_{\rm i} (1-\beta\mu_{\rm i}). 
\end{split}
\end{equation}
If the particle gyrates around the comoving magnetic field $\bm{ B'}$ before leaving the flow, its final energy and flight direction can be written as $E'_{\rm f}=E'_{\rm i}$ and $\mu'_{\rm f}\equiv p'_{{\rm f},x}/ E'_{\rm i}$, which in the laboratory frame become
\begin{equation}\label{eq:EfEi}
E_{\rm f}=\Gamma^2 E_{\rm i} (1-\beta\mu_{\rm i})(1+\beta\mu'_{\rm f}),\quad 
\mu_{\rm f}=\frac{\mu'_{\rm f}+\beta}{1+\beta\mu'_{\rm f}}.
\end{equation}
If $\mu_{\rm f}=\mu_{\rm i,}$ the particle energy is unchanged, but typically $\mu_{\rm f}\neq \mu_{\rm i}$, which implies $E_{\rm f}\simeq\Gamma^2 E_{\rm i}$, 
%Most of the particles (initially isotropic in the laboratory frame) that enter the flow and have time to achieve isotropy in the flow frame are eventually released with a $\sim\Gamma^2$ boost,
similar to a Compton scattering against a relativistic magnetic wall.
This phenomenon, which is independent of where particles enter/leave the flow, is well known for relativistic shocks: the energy gain is $\sim 2\Gamma^2$ in the first upstream--downstream--upstream cycle ($\mu_{\rm i}\simeq -1,\mu_{\rm f}\simeq 1$), but $\lesssim2$ in the following ones because particles are re-caught by the shock with $\mu\gtrsim 1-1/\Gamma\sim \mu_{\rm f}$ \citep{Achterberg+01}.

Let us consider a particle with $\gamma_{\rm in}\gg\Gamma$ entering the flow with $\mu_{\rm i}=0$, corresponding to $\mu'_{\rm i}=-\beta$, and assume $\bm{ B'}=-B'\bm{ z}$.
In the flow frame, the particle performs a Larmor gyration with frequency $\Omega'\equiv eB'/(\gamma_{\rm in}\Gamma mc)$ and in turn (Equation \ref{eq:EfEi}): 
\begin{equation}\label{eq:pi2}
\frac{E_{\rm f}}{E_{\rm i}}\simeq \Gamma^2 \left[1- \beta^2\cos{\Omega' t'}+\frac{\beta}{\Gamma}\sin{\Omega' t'}\right].
\end{equation}
The total energy gain depends on the phase $\varphi'\equiv \Omega't'\pmod{2\pi}$ when the particle leaves the flow: it is maximum ($2 \Gamma^2$) for $\varphi'_{\rm f}=\pi/2$, and $\sim \Gamma^2$ for $\varphi'_{\rm f}\in [\pi/2,3\pi/2]$. 
A boost of $\sim \Gamma^2$ in the laboratory requires the particle to stay in the flow for $T_{\rm acc}\gtrsim \Gamma\pi/(2\Omega')$, during which it travels a distance
\begin{equation}\label{eq:crit}
D_{\rm acc}\simeq T_{\rm acc} c\approx 4 {\rm kpc} \frac{\gamma_{\rm f}}{5\times 10^9 B'_{\mu{\rm G}}},
\end{equation}
with $\gamma_{\rm f}\equiv\gamma_{\rm in}\Gamma^2\sim 5\times 10^9$ the maximum UHECR Lorentz factor.
In reality, since relativistic flows diverge and the magnetic field drops \citep[$B'\sim$ few G $x_{\rm pc}^{-1}$ in blazar jets, e.g.,][]{sg09}, particles eventually escape, either because they reach the termination shock or because the condition in Equation \ref{eq:hillas} is violated.
Also, the expected radial dependence of the toroidal magnetic field induces an axial $\nabla \bm{ B}-$drift toward the flow head.

\begin{figure}%[htbp]
\begin{center}
\includegraphics[trim=0px 0px 0px 0px, clip=true, width=.45\textwidth]{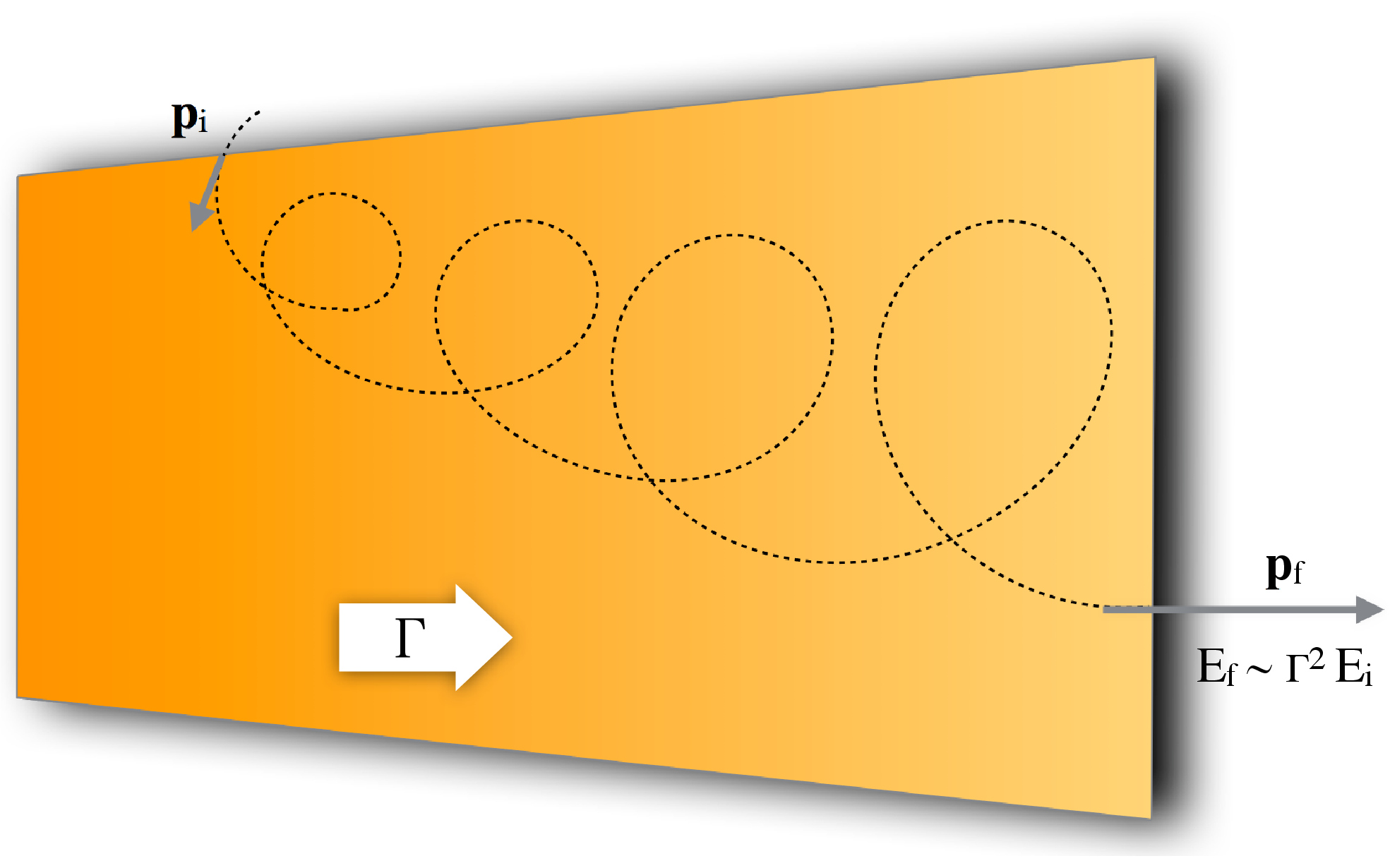}
\caption{Schematic trajectory of a galactic CR reaccelerated by a relativistic jet (not in scale). The total acceleration due to the motional electric field does not depend on the exact trajectory: a rotation $\gtrsim \pi/2$ around the jet magnetic field is sufficient to achieve a  $\sim\Gamma^2$ boost (Equation \ref{eq:pi2}).}
\label{fig:jet}
\end{center}
\end{figure}

In realistic flows, we can assume that the orbit is generally truncated with a random phase $\varphi'_{\rm f}$, which leads to an average energy gain $\langle E_{\rm f}/E_{\rm i}\rangle_{\varphi'_{\rm f}}=\Gamma^2$ (Equation \ref{eq:pi2}).
Figure \ref{fig:jet} shows the sketch of a possible particle trajectory in a conical (expanding) jet. 
Exact trajectories in realistic velocity/magnetic profiles of AGN jets will be presented in a forthcoming publication, but we anticipate that $\sim \Gamma^2$ energy gains are indeed common in astrophysical jets.

UHECR acceleration via such a one-shot (\emph{espresso}) mechanism thus requires either ultra-relativistic flows with $\Gamma\gtrsim 10^5$, or moderate Lorentz factors and pre-accelerated particles.
We now consider the case of AGN jets reaccelerating energetic CR seeds.

\subsection{``Seeds''} Galactic CRs accelerated in SNRs \citep{tycho,pionbump} represent natural seed candidates.
The maximum energy $E_{\rm max}\approx$ few $Z$ PeV (the CR ``knee'') is achieved before the SNR enters the Sedov stage ($t\approx t_{\rm S}$), when the shock velocity $V_{\rm S}$ starts to decrease because of the inertia of the swept-up material \citep{bac07}.
$E_{\rm max}$ can be estimated by equating $t_{\rm S}$ and the acceleration time $t_{\rm acc}$, which scale as
\begin{equation}
t_{\rm acc}\propto \frac{E_{\rm max}}{BV_{\rm S}^2};\quad
t_{\rm S}\propto\frac{1}{V_{\rm S}}\sqrt[3]{\frac{M_{\rm ej}}{\rho}};
\quad V_{\rm S}\approx\sqrt{\frac{2E_{\rm SN}}{M_{\rm ej}}},
\end{equation}
where $E_{\rm SN}$ and $M_{\rm ej}$ are the SN ejecta kinetic energy and mass, and $\rho$ and $B$ are the circumstellar density and magnetic field;
we have also used Bohm diffusion to derive $t_{\rm acc}$ \citep[see][for a justification of this assumption based on ab-initio simulations]{MFA,diffusion}.
Finally, we have
\begin{equation}
E_{\rm max} \propto Z B\rho^{-1/3} \propto Z\rho^{1/6}\sqrt{T_{\rm vir}},
\end{equation}
where we also assumed equipartition between thermal and magnetic pressures (as in the Milky Way), i.e., $B^2\propto \rho T$, with a typical temperature $T$ of the order of the virial temperature $T_{\rm vir}$.
Since the dependence on $\rho$ is very weak, and since $T_{\rm vir}$ is proportional to the total galactic mass, which does not differ greatly from galaxy to galaxy, $E_{\rm max}$ is expected to be roughly the same for any SNRs expanding in the interstellar medium. 
For core-collapse SNe, instead, $E_{\rm max}$ is achieved while the shock is still propagating in the wind launched in the pre-SN stages \citep{bell+13, cardillo+15}, which should be independent of the properties of the galaxy.
We conclude that the maximum energy of CRs accelerated in SNRs should be rather universal and correspond to the CR knee in the Milky Way.

\subsection{``Steam''}
Let us consider a galaxy that hosts a powerful active nucleus launching a relativistic jet with $\Gamma\approx 30$ and an opening angle of $\Delta\vartheta\approx 2^\circ$, which propagates for $H_{\rm j}\sim$ several kpc through diffuse galactic CRs.
The typical gyroradius of knee nuclei ($\sim 1{\rm pc}~ E_{\rm PeV}/Z B_{\mu{\rm G}}^{-1})$ is much smaller than the transverse size of the jet ($R_{\rm j}\approx H_{\rm j}\Delta\vartheta$), but larger than the jet boundary layer, whose thickness is determined by small-scale plasma processes.

We now estimate the fraction of galactic CRs that can percolate through the jet's lateral surface.
If $\dot{N}_{\rm CR}$ is the CR production rate in the galactic disk of radius $R_{\rm g}$, the CR flux in the halo reads $\Phi_{\rm CR}\simeq\dot{N}_{\rm CR}/(2\pi R_{\rm g}^2)$, and the number of CRs entering the two jets is $\dot{N}_{\rm j}\simeq \Phi_{\rm CR} 2 \pi \Delta\vartheta H_{\rm j}^2$.
Finally, the fraction of galactic CRs that can be espresso-accelerated is $\dot{N}_{\rm j}/\dot{N}_{\rm CR}\simeq 3.5\% (H_{\rm j}/R_{\rm g})^2(\Delta\vartheta/2^{\circ})$.
Extended jets with $H_{\rm j}\sim R_{\rm g}$, can boost a few percent of the knee nuclei by a factor of $\Gamma^2\approx 10^3$, producing UHECRs with energies beyond $5Z\times 10^{18}$\,eV, and in particular iron nuclei with energies $\gtrsim 10^{20}$\,eV.
Also CR electrons, whose Galactic spectrum is cut off around 1 TeV, may be reprocessed via the same mechanism; 
however, it is not obvious that even TeV electrons have gyroradii large enough to penetrate into the jet and, in general, radiative losses should prevent them from being accelerated to very high energies.

\subsection{Spectrum and Chemical composition}

\begin{figure}[htbp]
\begin{center}
\includegraphics[trim=30px 40px 30px 40px, clip=true, width=.5\textwidth]{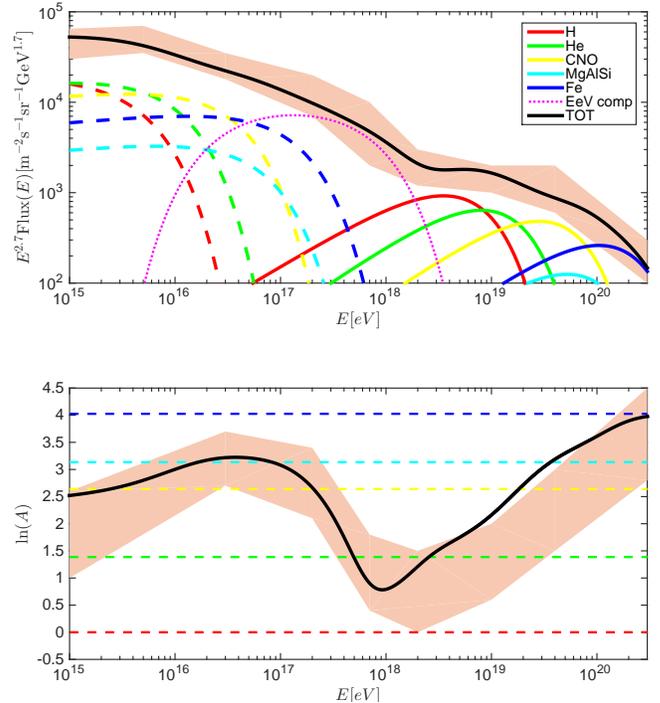}
\caption{Top panel: particle fluxes above $10^{15}$\,eV, for different species as in the legend. 
Dashed lines correspond to CRs accelerated in SNRs, and solid lines to UHECRs produced via espresso acceleration in AGN jets with $\Gamma\approx 30$.
The dotted line indicates the extra ``EeV component'' (see text).
Bottom panel: predicted average atomic mass $A$ as a function of energy; dashed lines correspond to the elements of the top panel.
Colored bands represent data from various experiments \citep[][and references therein]{KU12,GST13,Auger14a,Auger14b}.}
\label{fig:spec}
\end{center}
\end{figure}

A natural prediction of the proposed mechanism is that the chemical composition of galactic-like CRs, which is increasingly heavy above $10^{13}$\,eV and dominated by iron nuclei around $10^{17}$\,eV \cite[e.g.,][]{KU12}, should be mapped into UHECRs. 
This scenario is supported by recent Auger observations, which suggest a proton-dominated flux at $10^{18}$eV and a heavier composition at higher energies \citep{Auger14a}.
In particular, the composition is nitrogen-like at $\sim 4\times 10^{19}$\,eV, whereas the limited statistics do not yet allow conclusive measurements at higher energies; 
the inferred trend does not exclude an iron contribution above $10^{20}$\,eV. 
These results are not inconsistent with the Telescope Array's report of a pure proton composition if the different apertures, event selection cuts, and Monte Carlo models are taken into account \citep{Pierog13}. 
With adequate statistics, the Telescope Array should be able to distinguish between a proton-only and the Auger mixed composition \citep[see, e.g.,][]{PAO-TA15}.

Energy fluxes of Galactic CRs can be parameterized as
\begin{equation}\label{eq:fluxes}
\phi_s(E)= K_{s} \left( \frac{E}{10^{12}{\rm \,eV}}\right)^{-q_s}\exp{\left(-\frac{E}{Z_s\, 10^{15}{\rm\,eV}}\right)},
\end{equation}
where CR species ($s$=H, He, C/N/O, Mg/Al/Si, Fe) are grouped according to their (effective) charge $Z_s$=1, 2, 7, 13, 26 and atomic mass $A_s$=1, 4, 13, 27, 56; their abundances are tuned to the ones measured at $10^{12}$\,eV, namely $K_{\rm H}\approx$$0.15 {\rm m}^{-2}\,{\rm s}^{-1}\,{\rm sr}^{-1}$, $K_s/K_{\rm H}\approx{1,0.46,0.30, 0.07,0.14}$, and $q_{\rm H}\approx 2.7$ and  $q_{s \rm \neq H}\approx 2.6$ \cite[e.g.,][]{hoerandel03,nuclei,KU12}.
We use these simplified scalings, which capture the essential features of the most abundant species in Galactic CRs, as proxies for CR seeds in other galaxies as well.
Interestingly, the CR spectrum reprocessed by AGN jets may be \emph{flatter} than the one inside the galaxy.
SNRs inject CRs in the disk ($z=0$) with a spectrum $\phi_{\rm inj}(E)\propto E^{-\gamma}$, propagate in the halo (of thickness $H\approx 3-5$kpc) with a diffusion coefficient $D(E)\propto E^{\delta}$, where $\delta\approx 0.5$, and escape after a time $\tau_{\rm diff}\sim H^2/D(E)$.
The solution of the diffusion equation \citep[e.g.,][]{Lipari14} returns an equilibrium spectrum of $\phi_{\rm eq}\propto E^{-\gamma-\delta}\sim E^{-2.7}$ for $|z|<H$, and $\phi\propto D(E)\partial_z \phi_{\rm eq}|_{z=H}\propto E^{-\gamma}$ for $|z|>H$.
Moreover, if the host galaxy is very dense (such that spallation losses dominate over diffusive escape), one finds equilibrium spectra of $\phi_{\rm eq}\propto E^{-\gamma}$, and high-galactic-altitude spectra as flat as $\propto E^{-\gamma+\delta}$ \citep{cardillo+15}.
It is hence possible for extended jets to sweep up CR seeds with relatively flat spectra. 
Finally, energy-dependent percolation into the jet \citep[and possibly seed acceleration at the boundary layer: see][]{ostrowsky98,ostrowsky00} may lead to flatter injection spectra with a low-energy cut-off determined by the minimum energy for which CRs can enter the jet.

To calculate the spectra produced via espresso acceleration, we take galactic-like CRs with the  composition in Equation \ref{eq:fluxes} and with fiducial injection spectra $\phi_s(E)\propto E^{-2}$, and boost them in energy by a factor of $\Gamma^2\approx 10^3$.
Note that reproducing UHECR spectra and composition only requires the reacceleration of CRs with rigidities about one decade below the knee.
The overall UHECR normalization is chosen in order to reproduce the observed fluxes, and is consistent with the efficiency outlined above.
There is a rising consensus \cite[][]{GST13,abb14,Taylor14} that an additional light component must be relevant at $E\lesssim10^{18}$\,eV in order to explain UHECR spectra and composition; 
we consider this extra ``EeV component'' as well, modeled as in the paper by \cite{abb14}. 
 
Figure \ref{fig:spec} shows the obtained fluxes and chemical composition above $10^{15}$\,eV (top and bottom panels, respectively), compared with UHECR data. 
The model qualitatively reproduces the observed spectral features and the characteristic light--heavy--light--heavy modulation.
The agreement with observations is rather good, even if not remarkable above $\sim 10^{19}$\,eV.
A proper description of source distribution, injection spectra, nuclei photo-disintegration, and proton photo-pion production would be needed to explain the fine structure of UHECR spectra and composition \cite[][]{GST13,abb14,Taylor14}, but is beyond the scope of this Letter.
Nevertheless, we argue that inelastic losses should make the composition lighter at the highest energies, in better agreement with observations.

\section{Discussion and conclusions}
Most of the ingredients of the presented model are not completely new, but we believe the proposed model to be quite original in its entirety.
For instance, Ostrowski pointed out the possibility of generating UHECRs in fast blazer jets via \emph{diffusive} acceleration of thermal particles at the jet boundary \citep{ostrowsky90,ostrowsky98,ostrowsky00}. 
Instead, the espresso mechanism relies on energetic CR seeds and on a more general one-shot acceleration, which in addition returns the UHECR composition consistent with observations. 
Also the idea of producing UHECRs via CR reacceleration has already been suggested, either via relativistic matter plasmoids ejected by accreting black holes and neutron star, \citep[``cannonballs,'' see][]{cannonball08}, or via the interaction of AGN jets with large star-forming gaseous shells \cite[e.g.,][and references therein]{biermann+09,Gopal+10}.
However, such models need quite peculiar environmental structures whose cross-section is much smaller than the jet lateral surface that matters for espresso acceleration. 

The Lorentz factors inferred from multi-wavelength studies of blazars can be as high as 30--40 \citep{Tavecchio+10,Zhang+14}.
Some powerful sources  (e.g., Mrk 421 and Mrk 501\footnote{Quite intriguingly, Mrk 421 lies within the \emph{hot-spot} in the arrival direction of UHECRs above 57\,EeV reported by the Telescope Array \citep{TAhotspot}}, for which $L_{\rm bol}\gtrsim 10^{45}\ergs$ and redshift $z\approx 0.03$) are found within the UHECR ``horizon,'' i.e., their distance is less than or comparable to the mean free path for inelastic processes, such as photo-pion production for UHE protons and photo-disintegration for heavier nuclei \cite[see, e.g.,][]{Dermer07,abb14}.
Photo-disintegration around sources can in principle alter the UHECR chemical composition significantly \citep[e.g.][]{unger+15}, but it is not  expected to be important for the espresso mechanism, if most of the reacceleration occurs well beyond the AGN broad line region \citep[see, e.g.,][]{Dermer07}. 

The low level of anisotropy measured in the UHECR arrival directions \citep[][]{Auger10,TAhotspot} suggests that the horizon encompasses several sources, not necessarily associated with copious emission in other bands, such as $\gamma-$rays.
The census of potentially-relevant active galaxies may not be limited to the few powerful \emph{known} blazars inside the horizon: 
since some espresso-accelerated UHECRs may escape the jet sideways, also radio galaxies ---whose $\gamma$-ray emission is not beamed in our direction--- may partially contribute to the measured UHECR flux. 

In general, different types of AGNs (not just powerful BL Lac objects and flat-spectrum radio quasars) may be responsible for the global shape of the galactic to extragalactic transition.
The EeV component, for instance, may be produced via espresso acceleration in galaxies hosting jets with moderate luminosities and $\Gamma\lesssim 5$ (see Equation \ref{eq:Lcr}), which are numerous inside the horizon.
Such AGNs could be responsible for the production of the multi-PeV neutrinos detected by IceCube \citep{icecubepev}. 
A more detailed account for AGN zoology (distribution in luminosity, Lorentz factors, and redshift) is needed to further probe this scenario.

Finally, we check whether espresso acceleration may work also in the ultra-relativistic flows of other candidate sources of UHECRs.
Pulsar winds  have very large $\Gamma\gtrsim 10^5$, and live in environments rich in CR seeds accelerated at the surrounding SNR shocks.
However, for the Crab parameters ($\Gamma\approx 10^5$, $D\approx 0.1$pc, and $B'\approx B_{\rm PWN}/\Gamma$, where $B_{\rm PWN}\approx 100\mu$G is the nebular field), CRs cannot achieve isotropization within the wind (see Equation \ref{eq:crit}); 
such a requirement is unlikely to be fulfilled also in newly born millisecond pulsars.
For GRBs, instead, the limited extent of their jets makes it impossible to reaccelerate a sizable fraction of galactic CR seeds.

In summary, we presented a general mechanism for the acceleration of UHECRs in energetic AGN jets.
If the relativistic flow is sufficiently extended (Equations \ref{eq:hillas} and \ref{eq:crit}), seed CRs with gyroradii large enough to cross the jet boundary can be boosted by a factor of $\sim\Gamma^2$ in energy and be accelerated up to $\sim 10^{20}$\,eV.
Under reasonable assumptions, such a one-shot \emph{espresso} acceleration can account for both the spectrum and chemical composition of  UHECRs (Figure \ref{fig:spec}).
In forthcoming works we will probe this scenario by: testing CR injection and energy gain in realistic jet structures; accounting for AGN phenomenology; and including UHECR propagation and losses.

\begin{acknowledgments}
This research was partially supported by NASA (grant NNX14AQ34G to D.C.) and facilitated by the Max-Planck/Princeton Center for Plasma Physics. 
The author wishes to thank the anonymous referee and P. Blasi, L. Sironi, G. Rocha da Silva, Y. Gallant,  A. Spitkovsky, O. Bromberg, K. Murase, D. Giannios, M. Zaldarriaga, K. Parfrey, G. Ghisellini, S. Wykes, and M. Salvati for precious suggestions and stimulating discussions. 
\end{acknowledgments}

%\bibliography{UHECRs}

\begin{thebibliography}{46}
\expandafter\ifx\csname natexlab\endcsname\relax\def\natexlab#1{#1}\fi

\bibitem[{{Aab} {et~al.}(2014{\natexlab{a}}){Aab}, {Abreu}, {Aglietta}, {Ahn},
  {Al Samarai}, {Albuquerque}, {Allekotte}, {Allen}, {Allison}, {Almela}, \&
  et~al.}]{Auger14a}
{Aab}, A., {Abreu}, P., {Aglietta}, M., {et~al.} 2014{\natexlab{a}},
  \href{http://dx.doi.org/10.1103/PhysRevD.90.122005}{\prd, 90, 122005}

\bibitem[{{Aab} {et~al.}(2014{\natexlab{b}}){Aab}, {Abreu}, {Aglietta}, {Ahn},
  {Al Samarai}, {Albuquerque}, {Allekotte}, {Allen}, {Allison}, {Almela}, \&
  et~al.}]{Auger14b}
---. 2014{\natexlab{b}},
  \href{http://dx.doi.org/10.1103/PhysRevD.90.122006}{\prd, 90, 122006}

\bibitem[{Aartsen {et~al.}(2013)}]{icecubepev}
Aartsen, M.~G., Abbasi, R., Ackermann, M. {et~al.} 2013, 
\href{http://dx.doi.org/10.1103/PhysRevD.88.112008}{\prd, 88, 112008}

\bibitem[{{Abbasi} {et~al.}(2015){Abbasi}, {Bellido}, {Belz}, {de Souza},
  {Hanlon}, {Ikeda}, {Lundquist}, {Sokolsky}, {Stroman}, {Tameda}, {Tsunesada},
  {Unger}, {A.~Yushkov for the Pierre Auger Collaboration}, \& {the Telescope
  Array Collaboration}}]{PAO-TA15}
{Abbasi}, R., {Bellido}, J., {Belz}, J., {et~al.} 2015, ArXiv e-prints,
  \href{http://arxiv.org/abs/1503.07540}{{\sffamily arXiv:1503.07540
  [astro-ph.HE]}}

\bibitem[{{Abbasi et al.}(2014)}]{TAhotspot}
{Abbasi}, R.~U., Abe, M., Abu--Zayyad, T., {et~al.} 2014,
  \href{http://dx.doi.org/10.1088/2041-8205/790/2/L21}{\apjl, 790, L21}

\bibitem[{{Abreu} {et~al.}(2010){Abreu}, {Aglietta}, {Ahn}, {Allard},
  {Allekotte}, {Allen}, {Alvarez Castillo}, {Alvarez-Mu{\~n}iz}, {Ambrosio},
  {Aminaei}, \& et~al.}]{Auger10}
{Abreu}, P., {Aglietta}, M., {Ahn}, E.~J., {et~al.} 2010,
  \href{http://dx.doi.org/10.1016/j.astropartphys.2010.08.010}{\APh, 34, 314}

\bibitem[{{Achterberg} {et~al.}(2001){Achterberg}, {Gallant}, {Kirk}, \&
  {Guthmann}}]{Achterberg+01}
{Achterberg}, A., {Gallant}, Y.~A., {Kirk}, J.~G., \& {Guthmann}, A.~W. 2001,
  \href{http://dx.doi.org/10.1046/j.1365-8711.2001.04851.x}{MNRAS, 328, 393}

\bibitem[{{Ackermann et al.}(2013)}]{pionbump}
{Ackermann}, M.,  Ajello, M., Allafort, A., {et~al.} 2013,
  \href{http://dx.doi.org/10.1126/science.1231160}{Sci, 339, 807}

\bibitem[{{Aloisio} {et~al.}(2014){Aloisio}, {Berezinsky}, \& {Blasi}}]{abb14}
{Aloisio}, R., {Berezinsky}, V., \& {Blasi}, P. 2014,
  \href{http://dx.doi.org/10.1088/1475-7516/2014/10/020}{\jcap, 10, 20}

\bibitem[{{Bell} {et~al.}(2013){Bell}, {Schure}, {Reville}, \&
  {Giacinti}}]{bell+13}
{Bell}, A.~R., {Schure}, K.~M., {Reville}, B., \& {Giacinti}, G. 2013,
  \href{http://dx.doi.org/10.1093/mnras/stt179}{MNRAS, 431, 415}

\bibitem[{{Biermann} {et~al.}(2009){Biermann}, {Becker}, {Caramete}, {Curu{\c
  t}iu}, {Engel}, {Falcke}, {Gergely}, {Isar}, {Mari{\c s}}, {Meli}, {Kampert},
  {Stanev}, {Ta{\c s}c{\u a}u}, \& {Zier}}]{biermann+09}
{Biermann}, P.~L., {Becker}, J.~K., {Caramete}, L., {et~al.} 2009,
  \href{http://dx.doi.org/10.1016/j.nuclphysbps.2009.03.069}{NuPhS, 190, 61}

\bibitem[{{Blasi} {et~al.}(2007){Blasi}, {Amato}, \& {Caprioli}}]{bac07}
{Blasi}, P., {Amato}, E., \& {Caprioli}, D. 2007,
  \href{http://dx.doi.org/10.1111/j.1365-2966.2006.11412.x}{MNRAS, 375, 1471}

\bibitem[{{Blasi} {et~al.}(2000){Blasi}, {Epstein}, \& {Olinto}}]{Blasi+00}
{Blasi}, P., {Epstein}, R.~I., \& {Olinto}, A.~V. 2000,
  \href{http://dx.doi.org/10.1086/312626}{\apjl, 533, L123}

\bibitem[{{Caprioli} {et~al.}(2011){Caprioli}, {Blasi}, \& {Amato}}]{nuclei}
{Caprioli}, D., {Blasi}, P., \& {Amato}, E. 2011,
  \href{http://dx.doi.org/10.1016/j.astropartphys.2010.10.011}{APh, 34, 447}

\bibitem[{{Caprioli} \& {Spitkovsky}(2014{\natexlab{a}})}]{DSA}
{Caprioli}, D., \& {Spitkovsky}, A. 2014{\natexlab{a}},
  \href{http://dx.doi.org/10.1088/0004-637X/783/2/91}{\apj, 783, 91}

\bibitem[{{Caprioli} \& {Spitkovsky}(2014{\natexlab{b}})}]{MFA}
---. 2014{\natexlab{b}},
  \href{http://dx.doi.org/10.1088/0004-637X/794/1/46}{\apj, 794, 46}

\bibitem[{{Caprioli} \& {Spitkovsky}(2014{\natexlab{c}})}]{diffusion}
---. 2014{\natexlab{c}},
  \href{http://dx.doi.org/10.1088/0004-637X/794/1/47}{\apj, 794, 47}

\bibitem[{{Cardillo} {et~al.}(2015){Cardillo}, {Amato}, \&
  {Blasi}}]{cardillo+15}
{Cardillo}, M., {Amato}, E., \& {Blasi}, P. 2015,
  \href{http://dx.doi.org/10.1016/j.astropartphys.2015.03.002}{\APh, 69, 1}

\bibitem[{{Dar} \& {de R{\'u}jula}(2008)}]{cannonball08}
{Dar}, A., \& {de R{\'u}jula}, A. 2008,
  \href{http://dx.doi.org/10.1016/j.physrep.2008.05.004}{PhR, 466, 179}

\bibitem[{{Dermer}(2007)}]{Dermer07}
{Dermer}, C.~D. 2007, ArXiv e-prints,
  \href{http://arxiv.org/abs/0711.2804}{{\sffamily arXiv:0711.2804}}

\bibitem[{{Fang} {et~al.}(2012){Fang}, {Kotera}, \& {Olinto}}]{Fang+12}
{Fang}, K., {Kotera}, K., \& {Olinto}, A.~V. 2012,
  \href{http://dx.doi.org/10.1088/0004-637X/750/2/118}{\apj, 750, 118}

\bibitem[{{Gaisser} {et~al.}(2013){Gaisser}, {Stanev}, \& {Tilav}}]{GST13}
{Gaisser}, T.~K., {Stanev}, T., \& {Tilav}, S. 2013,
  \href{http://dx.doi.org/10.1007/s11467-013-0319-7}{FrPhy, 8,
  748}

\bibitem[{{Ghisellini} {et~al.}(2009){Ghisellini}, {Tavecchio}, \&
  {Ghirlanda}}]{GTG09}
{Ghisellini}, G., {Tavecchio}, F., \& {Ghirlanda}, G. 2009,
  \href{http://dx.doi.org/10.1111/j.1365-2966.2009.15397.x}{MNRAS, 399, 2041}

\bibitem[{{Gopal-Krishna} {et~al.}(2010){Gopal-Krishna}, {Biermann}, {de
  Souza}, \& {Wiita}}]{Gopal+10}
{Gopal-Krishna}, {Biermann}, P.~L., {de Souza}, V., \& {Wiita}, P.~J. 2010,
  \href{http://dx.doi.org/10.1088/2041-8205/720/2/L155}{\apjl, 720, L155}

\bibitem[{{Hillas}(1984)}]{hillas84}
{Hillas}, A.~M. 1984,
  \href{http://dx.doi.org/10.1146/annurev.aa.22.090184.002233}{ARA\&A, 22, 425}

\bibitem[{{H{\"o}randel}(2003)}]{hoerandel03}
{H{\"o}randel}, J.~R. 2003,
  \href{http://adsabs.harvard.edu/abs/2003APh....19..193H}{APh, 19, 193}

\bibitem[{{Jiang} {et~al.}(2007){Jiang}, {Fan}, {Ivezi{\'c}}, {Richards},
  {Schneider}, {Strauss}, \& {Kelly}}]{Jiang+07}
{Jiang}, L., {Fan}, X., {Ivezi{\'c}}, {\v Z}., {et~al.} 2007,
  \href{http://dx.doi.org/10.1086/510831}{\apj, 656, 680}

\bibitem[{{Kampert} \& {Unger}(2012)}]{KU12}
{Kampert}, K.-H., \& {Unger}, M. 2012,
  \href{http://dx.doi.org/10.1016/j.astropartphys.2012.02.004}{\APh, 35, 660}

\bibitem[{{Katz} {et~al.}(2013){Katz}, {Waxman}, {Thompson}, \&
  {Loeb}}]{Katz+13}
{Katz}, B., {Waxman}, E., {Thompson}, T., \& {Loeb}, A. 2013, ArXiv e-prints,
  \href{http://arxiv.org/abs/1311.0287}{{\sffamily arXiv:1311.0287
  [astro-ph.HE]}}

\bibitem[{{Lipari}(2014)}]{Lipari14}
{Lipari}, P. 2014, ArXiv e-prints,
  \href{http://arxiv.org/abs/1407.5223}{{\sffamily arXiv:1407.5223
  [astro-ph.HE]}}

\bibitem[{{Morlino} \& {Caprioli}(2012)}]{tycho}
{Morlino}, G., \& {Caprioli}, D. 2012,
  \href{http://dx.doi.org/10.1051/0004-6361/201117855}{A\&A, 538, A81}

\bibitem[{{Murase} {et~al.}(2012){Murase}, {Dermer}, {Takami}, \&
  {Migliori}}]{Murase+12}
{Murase}, K., {Dermer}, C.~D., {Takami}, H., \& {Migliori}, G. 2012,
  \href{http://dx.doi.org/10.1088/0004-637X/749/1/63}{\apj, 749, 63}

\bibitem[{{Ostrowski}(1990)}]{ostrowsky90}
{Ostrowski}, M. 1990, A\&A, 238, 435

\bibitem[{{Ostrowski}(1998)}]{ostrowsky98}
---. 1998, A\&A, 335, 134

\bibitem[{{Ostrowski}(2000)}]{ostrowsky00}
---. 2000, \href{http://dx.doi.org/10.1046/j.1365-8711.2000.03146.x}{MNRAS,
  312, 579}

\bibitem[{{O'Sullivan} \& {Gabuzda}(2009)}]{sg09}
{O'Sullivan}, S.~P., \& {Gabuzda}, D.~C. 2009,
  \href{http://dx.doi.org/10.1111/j.1365-2966.2009.15428.x}{MNRAS, 400, 26}

\bibitem[{{Pierog}(2013)}]{Pierog13}
{Pierog}, T. 2013,
  \href{http://dx.doi.org/10.1088/1742-6596/409/1/012008}{JPhCS, 409, 012008}

\bibitem[{{Sironi} \& {Spitkovsky}(2011)}]{ss11}
{Sironi}, L., \& {Spitkovsky}, A. 2011,
  \href{http://dx.doi.org/10.1088/0004-637X/726/2/75}{\apj, 726, 75}

\bibitem[{{Tavecchio} {et~al.}(2010){Tavecchio}, {Ghisellini}, {Ghirlanda},
  {Foschini}, \& {Maraschi}}]{Tavecchio+10}
{Tavecchio}, F., {Ghisellini}, G., {Ghirlanda}, G., {Foschini}, L., \&
  {Maraschi}, L. 2010,
  \href{http://dx.doi.org/10.1111/j.1365-2966.2009.15784.x}{MNRAS, 401, 1570}

\bibitem[{{Taylor}(2014)}]{Taylor14}
{Taylor}, A.~M. 2014,
  \href{http://dx.doi.org/10.1016/j.astropartphys.2013.11.006}{\APh, 54, 48}

\bibitem[{{Unger} {et~al.}(2015){Unger}, {Farrar}, \& {Anchordoqui}}]{unger+15}
{Unger}, M., {Farrar}, G.~R., \& {Anchordoqui}, L.~A. 2015, ArXiv e-prints,
  \href{http://arxiv.org/abs/1505.02153}{{\sffamily arXiv:1505.02153
  [astro-ph.HE]}}

\bibitem[{{Vietri}(1995)}]{vietri95}
{Vietri}, M. 1995, \href{http://dx.doi.org/10.1086/176448}{\apj, 453, 883}

\bibitem[{{Waxman}(1995)}]{waxman95}
{Waxman}, E. 1995, \href{http://dx.doi.org/10.1103/PhysRevLett.75.386}{PRL, 75, 386}

\bibitem[{{Waxman}(2004)}]{Waxman04}
---. 2004, \href{http://dx.doi.org/10.1088/1367-2630/6/1/140}{NJPh, 6, 140}

\bibitem[{{Woo} \& {Urry}(2002)}]{woo-urry02}
{Woo}, J.-H., \& {Urry}, C.~M. 2002,
  \href{http://dx.doi.org/10.1086/342878}{\apj, 579, 530}

\bibitem[{{Zhang} {et~al.}(2014){Zhang}, {Zhao}, \& {Cao}}]{Zhang+14}
{Zhang}, B., {Zhao}, X., \& {Cao}, Z. 2014,
  \href{http://dx.doi.org/10.4236/ijaa.2014.43046}{IJAA, 4, 499}

\end{thebibliography}
%\bibliographystyle{yahapj}

\end{document}